\documentclass{article}
\usepackage{color}
\usepackage{amssymb}
\usepackage{amsmath}
\usepackage{latexsym}
\usepackage{slashed, amsthm}
\usepackage{graphicx}

\title{The Counterfactual Photoelectric Effect}
\author{A. Ludu$^1$,  J. A. Morris$^{2}$, C. Rugina$^3$
	\\
	\\ \small{$^1$ Department of Mathematics, Embry-Riddle Aeronautical University, Daytona Beach, FL, USA}
	\\ \small{$^2$ Alumnus, Department of Physics, Northeastern University, Boston MA, USA}
	\\ \small{$^3$ Department of Physics, University of Bucharest, Magurele, Romania}
	\\}

\begin{document}

\maketitle 
{\let\thefootnote\relax\footnotetext{{\em Emails}: \\ christina.rugina11@alumni.imperial.ac.uk, ludua@erau.edu, josh.a.morris.1@gmail.com}}

\begin{abstract}

\noindent 

We explore a counterfactual protocol for energy transfer.  A modified version of a Mach-Zehnder interferometer dissociates a photon's position and energy into separate channels, resulting in a photoelectric effect in one channel without the absorption of a photon.  We use the quantum Zeno effect to extend our results by recycling the same photon through the system and obtain a stream of photoelectrons.  If dissociation of properties such as energy can be demonstrated experimentally, there may be a variety of novel energy-related applications that may arise from the capacity to do non-local work.  The dissociation of intrinsic properties, like energy, from elementary particles may also lead to theoretical discussions of the constitution of quantum objects.

\end{abstract}

\vskip0.7cm

\vskip0.7cm

\section*{Introduction}

\noindent 
The productive interplay of science and technology is nowhere more on display than in cutting-edge research on the foundations of quantum mechanics and the development of quantum technologies.  Here we focus on research in counterfactual quantum protocols, which achieve transformative effects in the absence of local processes.  Researchers have developed protocols that obtain the gedanken experiment output of a quantum computer that has not been run \cite{Zeno2}, images have been obtained without photons interacting with their subjects \cite{Image}, communication systems have been designed that can operate without sending or receiving any particles \cite{Comm}, and work at a distance has recently been proposed \cite{WorkAtADistance}.  Some theorists are beginning to speculate that counterfactuals may play a role in fundamental physics  \cite{AEEC,DDCT}, which may lead to new principles of physics and additional technological advances.  

\medskip
\noindent 
A large number of the counterfactual protocols that have been developed thus far have focused on information acquisition  \cite{Zeno2, Image, Comm}.  The prototypical work of Elitzur and Vaidman \cite{EVB} focused on interaction-free measurements to sort active bombs from duds, without detonating them.  While the physical processes of the counterfactual measurements in their experiment remained in the counterfactual realm, a number of counterfactual protocols have been subsequently developed that transform physical quantities and/or lead to physical resultants, and some of these have already been tested in the lab.  The counterfactual protocol developed here follows this line of research.

\medskip
\noindent  
Utilizing a modified Mach-Zehnder interferometer (MZI), a single-photon source, and specific pre- and post-selection stages, we demonstrate a method for guiding the dissociated energy of a photon to a cavity where a photoelectric interaction occurs without the absorption of a photon.  Extending the technique, the interaction can be repeated with a single photon in an arbitrary number of repeated cycles, producing a stream of photoelectrons.  The separation technique developed in $\cite{Aharonov1}$, dubbed the ``Quantum Cheshire Cat effect" (QCC), while controversial \cite{QCC, QCCCrit1, QCCCrit2}, has recently led to the experimental demonstration of the dissociation of neutrons from their magnetic moments \cite{Neu}.

\begin{figure}[h] 
   \centering
   \includegraphics[width=3.8in]{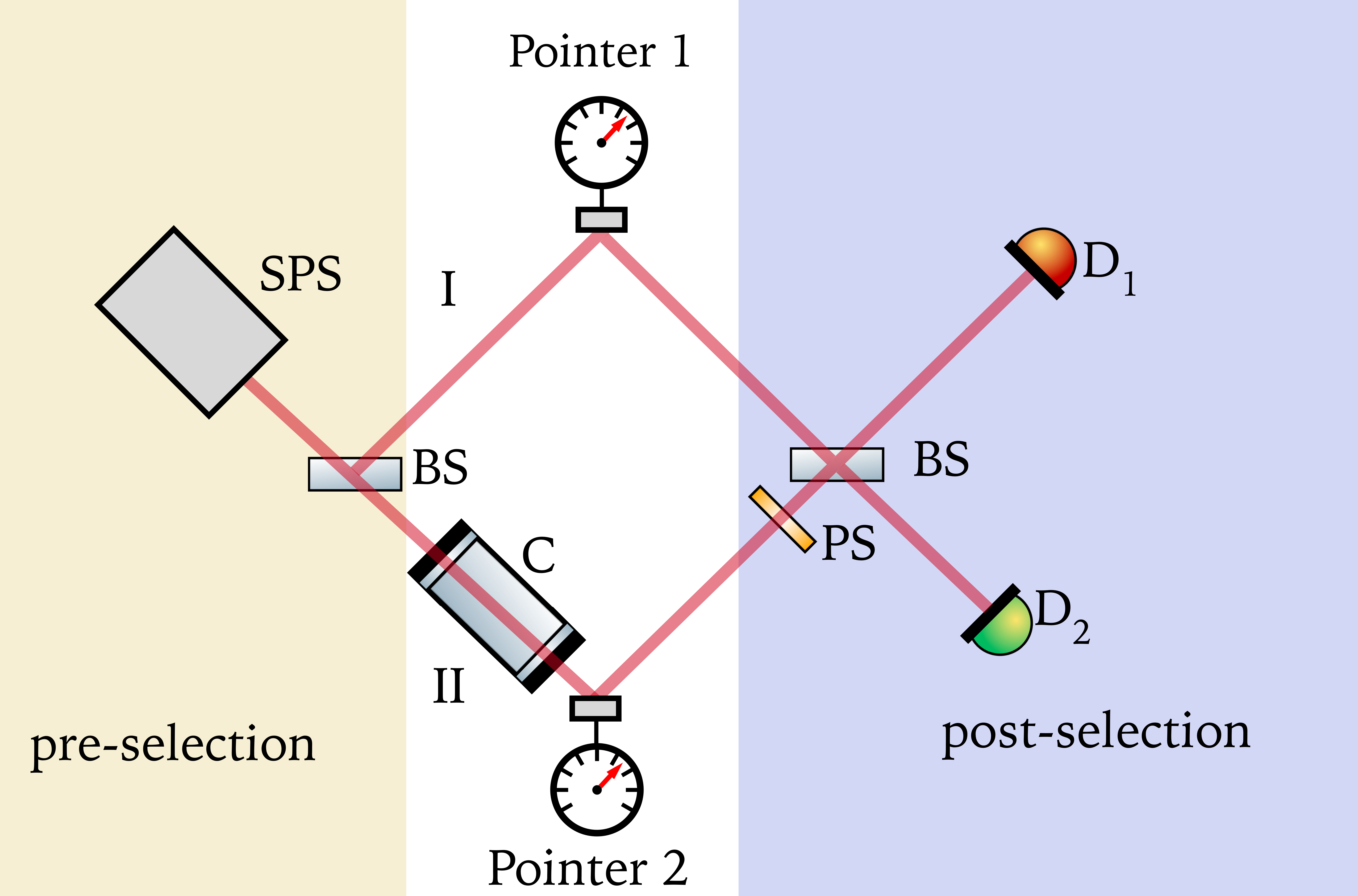} 
   \caption{Design of the experiment: A photon travels through arm I of the modified MZI, while its energy travels through arm II and collides with atoms in the cavity, C.  The experiment is counterfactual because no particle participates directly in the collision, yet an electron is still produced as in the normal effect.}  
   \label{fig:example}
\end{figure}

\section{The Counterfactual Photoelectric Effect}

\noindent 
We employ an MZI, as shown in figure 1.  A single photon source emits a photon in the pre-selection area with a wave function given by equation (1). Figure 1 contains: a photon source (SPS), two beam-splitters (BS), a phase-shifter (PS), a cavity for the interaction, two weak values measurements pointers, and two detectors, D\textsubscript{1} and D\textsubscript{2}. Once the photon reaches the post-selection stage, it is detected by detector D\textsubscript{1}.  In the experiment, the photon travels through arm I, while its dissociated energy travels through arm II.

\medskip
\noindent
The variant of the photoelectric effect demonstrated here is possible due to techniques of particle-property dissociation, as established in $\cite{Aharonov1}$. These techniques make it possible to  spatially dissociate a photon from its energy, and in our case, use the energy to induce a photoelectric effect while the carrier photon remains remote and is not absorbed in the process.

\medskip

\noindent 
The quantum dissociation technique is predicted by the Weak Values Framework (WVF), as discussed in $\cite{Aharonov1}$.  This technique allows for the spatial separation of a particle's property from the particle's location, namely, the property of a particle is present in a location where the particle is absent, and vice versa.   We know that weak values are determined by the pre-selected and post-selected wave functions, as boundary conditions and these determine in effect the separation of the particle from its property. The pre- and post-selected wave functions are

\begin{equation}
|\ pre\ \rangle \ = |\ 0\ \rangle \ |\ I\ \rangle  +\ \ ( |\ 0\ \rangle  +\ |\ 1\ \rangle )|\ II\ \rangle 
\end{equation}

\begin{equation}
|\ post\ \rangle\ = |\ 0\ \rangle |\ I\ \rangle +\  \ (|\ 0\ \rangle -\ |\ 1\ \rangle) |\ II\ \rangle 
\end{equation}

\medskip

\noindent 
Here, $|\ I\ \rangle$ and $|\ II\ \rangle$ indicate the presence of the photon in arms I or II, respectively, and $|\ 0\ \rangle$ and $|\ 1\ \rangle$ are the eigenstates of the number operator, with eigenvalues of 0 or 1. With these boundary conditions defined, it is clear that the photon is always present in arm I.

\medskip

\noindent 
Following $\cite{Aharonov1, QCC}$, we note that our results hold in the WVF, and that the QCC in our case is realized with the dissociation of position and energy. What we will show is that we end up with an absorption of energy in this effect, and that the effect is not corpuscular, as in the ordinary photoelectric effect, but wave-like. We can also characterize the dissociation here as being between the wave and the particle aspects of the photon.  Since the photon's energy wave is present in arm II, and the photon itself stays in arm I, the energy alone causes the photoelectric effect in this case, while the photon does not play a direct role as a particle, and is retained in the post-selection stage for re-cycling. This unusual process is the counterfactual photoelectric effect. 

\medskip
\noindent
Using the technique developed in $\cite{Aharonov1, QCC}$, we present the mathematics of the QCC effect for CF PE below. To begin with, the Hamiltonian of electromagnetic field associated the photon is

\begin{equation}
H= (a^\dagger a + \frac{1}{2})
\end{equation}

\noindent
So the above equation is momentarily and weakly measured at a certain point in time (rather than governing the whole time evolution).
We define the weak value of an operator A as:

\begin{equation}
A^w = \frac{\langle \ pre \ | \ A \ | \ post \ \rangle}{\langle \ pre \ | \ post\ \rangle  }
\end{equation}

\noindent
The QCC effect is formulated within the WVF, and therefore we work with the von Neumann coupling: $exp(-ig H^w_I P_I)$ where index I stands for arm I. We use the von Neumann coupling to verify what the weak value is, we are not yet talking about the photoelectric effect. Note that we use here the Hamiltonian as the operator we measure at some intermediate point, rather than the Hermitian operator which describes the time evolution of the system throughout all times.

\medskip
\noindent
Now, in arm I, with the given boundary conditions:

\medskip
\noindent
$\Pi^w_I = 1$,  $ H^w_I=\frac{1}{2}$.

\medskip
\noindent
These values imply that a particle is present (in arm I) and some of its energy is also present, corresponding in our case to the photon being present in arm I, while its energy property is only partly present.

\medskip
\noindent
And for arm II:

\medskip
\noindent
 $\Pi^w_{II} = 0$,  $ H^w_{II}=1$.

\medskip
\noindent
These values tell us that the particle is not present in arm II. This corresponds to the case where the photon is absent, but the photon's energy property is present, which means the counterfactual photoelectric effect can occur. Usually, people identify a photon as energy and therefore they can say it exists in arm II as well as in arm I. We use a slightly different definition of the existence which corresponds only to projection operators.

\medskip
\noindent
Using the language and the techniques in $\cite{QCC}$, we have demonstrated that we have a QCC-like effect taking place in our setup, with the dissociated property of the photon being its energy and we stress that this is true in the weak value framework only and g is ntaurally small and the linearization of the von Neumann coupling exponent.  This energy allows for a counterfactual photoelectric effect in arm II, with a high probability.  When the QCC effect kicks in and the energy is absorbed by the atoms in the cavity, a `free` electron is released, as we demonstrate in the next section, while the photon itself survives absorption and is retrieved in the post-selection area.

\medskip
\noindent
Now, the wave function in the pre-selected area and the post-selected area is phase-shifted, according to $\cite{QCC}$, as

\begin{equation}
|\ \phi_I(t_f)\ \rangle\ = \frac{1}{2} exp(-ig H^w_I P_I) |\ pre\ \rangle
\end{equation}

\medskip
\noindent
Given that our set up is governed by relativistic quantum mechanics, we could conclude that our wave function's symmetry is described by the Lorentz group. However, our case shares some similarities, too, with the non-relativistic Bargmann group, the central extension of the Galilei group (no wonder, since the photoelectric effect can have a non-relativistic description, too). The phase shift of our wave function described by the equation above, suggests that there is an analogy with the Bargmann group and if that is the case, in principle this could be seen also as a 5-dimensional physical set-up and we see the extra dimension as a null coordinate of the Eisenhart lift in the phase shift (that is the 5-dimensional spacetime is described by the 4 regular spacetime coordinates and the fifth is the phase shift of the wave function).

\medskip

\noindent The phase shift is produced by the QCC effect, as it is a modified version of the Aharanov-Bohm effect.  This phase shift alone suggests an interaction occurred in arm II, even in the absence of our direct knowledge of an interaction between the energy and the atoms in the cavity. From the perspective of the Aharonov-Bohm effect, there is no background electromagnetic field, yet the photon's potential induces a phase shift in the wavefunction, which we observe independently of the outcome of the interaction.

\section{Recycling the Photon with the Zeno Effect}

\begin{figure}[h] 
   \centering
   \includegraphics[width=3in]{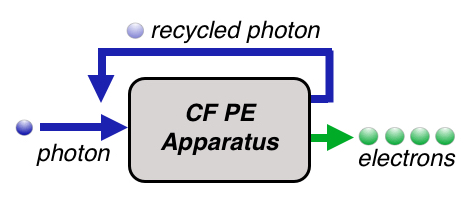} 
   \caption{The quantum Zeno effect can be used to extend our result such that a single, re-cycled photon, can continuously produce a photoelectric current.}
   \label{fig:example}
\end{figure}

\noindent
In this section we will see how a counterfactual current of free electrons at the end of the cavity placed in arm II of the MZI is formed, as a result of sending the photon in a Zeno cycle $\cite{Zeno}$. We depict the cavity in fig. 2. The left end of the cavity is in arm II of the MZ interferometer and the right end hosts a bulk of atoms and a wall permeable to the emitted electrons as a result of the photoelectric effect. The Zeno effect occurs when we are sending the same photon many times in a cycle - see picture above- and as in the case within the interaction-free measurements, the probability of `explosion` in each run is negligibly small; the photon is circulating in the MZI from the pre-selection area to the post-selection one, where it is picked up and fed back into the pre-selection area. This Zeno effect is responsible for the jet of free electrons we almost miraculously get and renders the whole process counterfactual.The Zeno effect also means that the probability of success in each stage is close to 1, and that is the case here as well.

\medskip
\noindent 
We start by remembering that in arm II we have no photon, just its energy, and that means that the wave function of that lump of energy at the left end of the cavity is given by:

\begin{equation}
|\ L\ \rangle\ = |\ p \epsilon\ \rangle |\ 0\ \rangle
\end{equation}

\medskip
\noindent
The zero occupation photon number state indicates that there is no photon at the left end, just momentum p and polarization $\epsilon$.

\medskip
\noindent
The interaction Hamiltonian for the photoelectric effect in a cavity of length L and polarization $\epsilon$ of the photon is:

\begin{equation}
H_{int} = \sqrt{\frac{2 \pi}{L^3 \omega}} [ a e ^{i\vec{p} \cdot \vec{r}} + a^\dagger e^{-i \vec{p} \cdot \vec{r}}] \vec{\epsilon} \cdot \vec{p}
\end{equation}

\medskip
\noindent
The interaction Hamiltonian for the electron emitted by an atom as a result of the photoelectric effect is:

\begin{equation}
H^e_{int} = \sqrt{\frac{2 \pi}{L^3 \omega}} [ a_e e ^{i\vec{p_e} \cdot \vec{r}} + {a^\dagger}_e e^{-i \vec{p_e} \cdot \vec{r}}] \vec{\epsilon_e} \cdot \vec{p_e}
\end{equation}  

\medskip
\noindent
We notice that the wave function $|\ L\ \rangle$ of the energy lump becomes at the right end of the cavity:

\begin{equation}
|\ R\ \rangle\ = |\ p \epsilon\ \rangle |\ 1\ \rangle
\end{equation}

\noindent
since 

\begin{equation}
|\ R\ \rangle\ = H_{int}  |L\rangle.
\end{equation}

\begin{figure}[h] 
   \centering
   \includegraphics[width=4in]{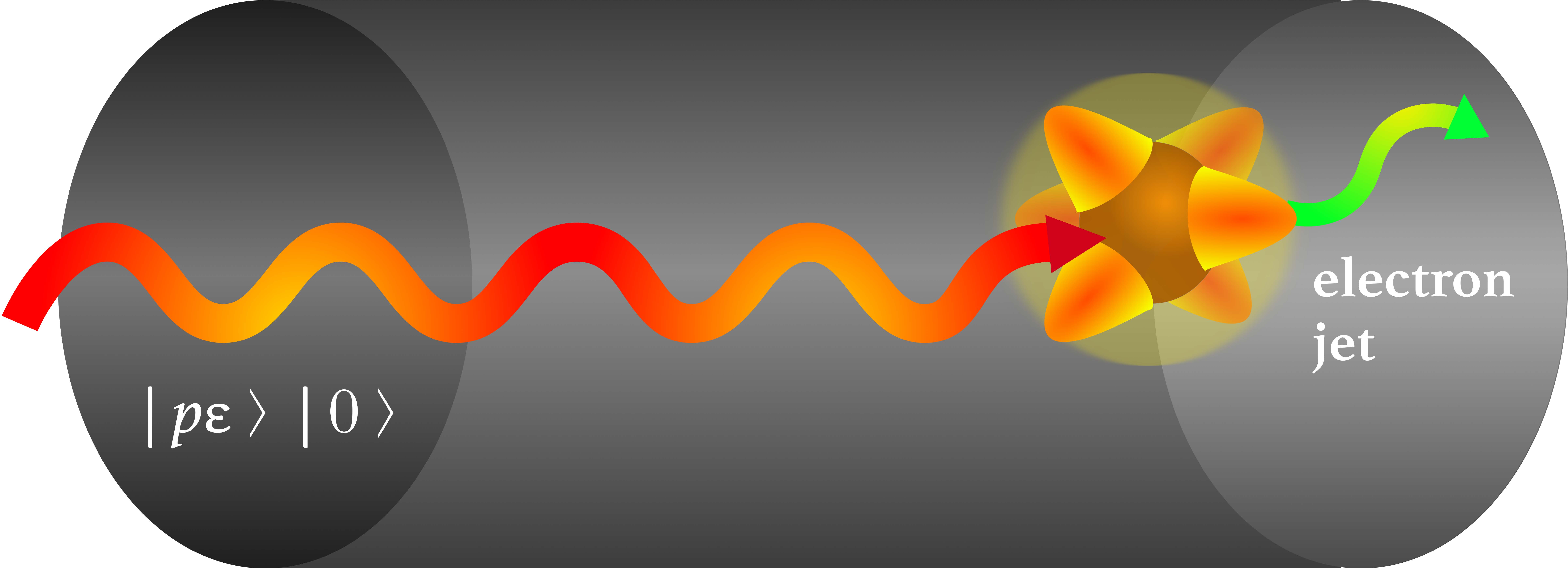} 
   \caption{The energy wave function--with zero occupation number--is present in the cavity and collides with a representative atom to produce a stream of electrons.}
   \label{fig:example}
\end{figure}

\medskip
\noindent
This means that the photon is regenerated at the right end and also the electron is emitted, due to the counterfactual photoelectric effect. Due to another effect described in $\cite{disapp}$ we will actually retrieve the photon in the post-selection area of the interferometer, which indicates a strong entanglement between the right end of the cavity and the post-selection area. This is possible as in the disappearing particle experiment, since the photon is retrieved in the post-selection area. The electron is emitted, too, as the electron interaction Hamiltonian surely indicates. 

\medskip
\noindent
This set-up and resolution indicates that with one photon sent to perform an N-cycle Zeno through the MZ interferometer, one can obtain a current of N free electrons, employing the Quantum Cheshire Cat effect and also the disappearing particle effect.

\section{Disappearing and Reappearing Particle Effect in Arm II}

\medskip
\noindent 
The absence of the photon in arm II, and the emergence of a photon in the post-selection stage, can be understood as an instance of the theoretical work presented in $\cite{disapp}$.  In that work, a particle apparently disappears from where it could have been expected to be found with certainty, in either Box A (at $t_{1}$ or Box B (at $t_{2}$), only to re-appear in a place where it could not be expected at all (Box C at $t_{3}$).  The key insight for this paradoxical thought experiment, is that the particle of interest does not actually disappear from anywhere at all, rather it did not really reside in either box to begin with.  The pre- and post- selected states are such that a particle only appears in Boxes A or B if a measurement is made at either $t_{1}$ or $t_{2}$, however without measurement, the states are orthogonal in the subspace spanned by the two boxes: no particle is actually present in either of them.  Further discussion of how to understand the contents of the boxes can be found in $\cite{weak}$, utilizing pairs of self-canceling weak values dubbed ``particle and nega-particle".  In our work, the ``disappearance" occurs in arm II and the ``reappearance" occurs in post-selection. The pre- and post-selected states are orthogonal, as in the theoretical work mentioned above, so one cannot find a particle along the second path if a measurement is made there. However, a pair of self-canceling weak values can be found, such as those of the operators $\Pi_{II}\Pi_0$ and $\Pi_{II} \Pi_1$, where $\Pi_0$ and $\Pi_1$ project on $|0\rangle$ and $|1\rangle$, respectively.  

\medskip
\noindent
We proved in the two vector state formalism the separation of energy from the location of the particle in accordance with predictions of $\cite{Aharonov1, disapp}$. In $\cite{Aharonov1}$ where separation of spin from location is achieved, separation of energy from location of a particle is explicitly mentioned as future direction of research. In $\cite{disapp}$ a disappearance/reappearance cycle is achieved and we mirror that in our own Zeno cycle and in that paper there is reached a separation from location of a non-local property, modular momentum. We extend that gedanken experiment here to include separation of energy.

\section*{Conclusion}

\noindent %
We have presented a counterfactual variant of the photoelectric effect, one that not only occurs without the absorption of a photon, but that produces a current of photoelectrons from a single photon in a Zeno cycle through a standard MZI. The effect can be described as a result of the Photoelectric effect, the Quantum Cheshire Cat effect, the Disappearing/Re-appearing particle effect, and the quantum Zeno effect. 

\medskip
\noindent 
Our work follows developments from $\cite{QCC}$, where a theoretical prediction is made about dissociating a photon's polarization from its location.  In recent experimental work $\cite{Neu}$, neutron magnetic moments are dissociated from their host neutrons.  Energy dissociation, as in the effect presented here, is interesting to theorists because intrinsic properties have been thought to be fundamental to particle constitution and identity, in a way that photon polarization or neutron magnetic moment are not.  What does the decoupling of a fundamental property mean for the definition of quantum objects?  Is a photon still considered a photon when it is separated from its energy?  Does the photon behave differently in this strange context?  These questions are still unanswered, but provide food for thought for future work.

\section*{Acknowledgements}
	The authors thank Yakir Aharonov, Virgil Baran, Gary Gibbons, Dana Ioan, Aurelian Isar, John Swain and Barton Zwiebach for reading the manuscript, making useful comments, and providing inspiration. Special thanks are due to Eli Cohen for support and guidance, without which this paper wouldn't be possible.


\begin{thebibliography}{99}
	
\bibitem{Zeno2}
O. Hosten, M. T. Rakher, J.T. Barreiro, N.A. Peters, P.G. Kwiat, ``Counterfactual quantum computation through quantum interrogation", \textit{Nature} {\bf 439}, 949 (2006)

\bibitem{Image}
A. White, J. Mitchell, O. Nairz, P. Kwiat, ``Interaction-free imaging", \textit{Physical Review A} {\bf 58}, 605 (1998)

\bibitem{Comm} 
J.W. Pan, et al., ``Direct counterfactual communication via quantum Zeno effect", \textit{PNAS} {\bf 114:19} 4920-4924 (2017)

\bibitem{WorkAtADistance}
C. Elouard, M. Waegell, B. Huard, A. N. Jordan, "Spooky work at a distance: an Interaction-Free quantum measurement-driven energy", arXiv: 1904.09289

\bibitem{AEEC} 
A. C. Elitzur and E. Cohen, ``1 - 1 = Counterfactual: on the potency and significance of quantum non-events", \textit{Phil. Trans. R. Soc. A}  {\bf 374:20150242} 1 (2016)


\bibitem{DDCT} 
D. Deutsch, C. Marletto, ``Constructor Theory of Information", \textit{Proc. R. Soc. A}  {\bf 471, 2174} 1 (2015)

\bibitem{EVB}
A. Elitzur, L. Vaidman, ``Quantum mechanical interaction-free measurements", \textit{Foundations of Physics} {\bf 23:7} 987--997 (1993)
	
\bibitem{Aharonov1}
Y. Aharonov, S. Popescu, D. Rohrlich, P. Skrzypczyk, ``Quantum Cheshire Cats", \textit{New J. Phys.} {\bf 15} 113015 (2013)

\bibitem{QCC}
Q. Duprey, S. Kanjilal, U. Sinha, D. Home, A. Matzkin, ``The Quantum Cheshire Cat effect: Theoretical basis and observational implications", \textit{Ann. Phys} {\bf 391} 1 (2018)

\bibitem {QCCCrit1} D. Sokolovski, ``Path probabilities for consecutive measurements, and certain `quantum paradoxes'\ ", \textit{Annals of Physics} (2018) https://doi.org/10.1016/j.aop.2018.05.017 

\bibitem {QCCCrit2} B. E. Y. Svensson, ``Non-representative Quantum Mechanical Weak Values", \textit{Found Phys} (2015) DOI 10.1007/s10701-015-9951-0 

\bibitem{Neu} 
T. Denkmayr, et al., ``Observation of a quantum Cheshire Cat in a matter-wave interferometer experiment", \textit{Nature Communications} {\bf 5:4492}, 1 (2014)

\bibitem{Zeno}
Y. Aharonov, E. Cohen, S. Popescu, "A current of the Cheshire Cat's smile: Dynamical Analysis of Weak Values", under revision, arXiv:1510.03087.

\bibitem{disapp}
Y. Aharonov, E. Cohen, A. Landau, A. C. Elitzur, "The case of the Disappearing (and Re-Appearing) Particle", \textit{Nature Scientific Reports} {\bf 7:531} (2017)

\bibitem{weak}
Y. Aharonov, E. Cohen, M. Waegell, and A. Elitzur, "The Weak Reality That Makes Quantum Phenomena More Natural: Novel Insights and Experiments", \textit{Entropy} {\bf 20(11)}, 854 (2018)




\end{thebibliography}
\end{document}